\documentclass[twocolumn,twoside,preprintnumbers,amsmath,amssymb]{revtex4}
\usepackage{epsfig}
\usepackage{graphicx}
\usepackage{dcolumn}
\usepackage{bm}

\newcommand*{\no}{\noindent}
\newcommand*{\bea}{\begin{eqnarray}}
\newcommand*{\eea}{\end{eqnarray}}
\newcommand*{\be}{\begin{equation}}
\newcommand*{\ee}{\end{equation}}
\newcommand*{\pd}{\partial}

\newcommand*{\pref}[1]{(\ref{#1})}

\newcommand*{\mn}{{\mu\nu}}

\newcommand*{\nn}{\nonumber}

\newcommand*{\indexsep}{,}
\newcommand*{\tl}{\mathrm{tl}}

\usepackage{fancyhdr}

\usepackage{pslatex}

\begin{document}

\title{{\Large On the infrared behavior of Green's functions in Yang-Mills theory}}

\author{Axel Maas}
\author{Attilio Cucchieri}
\author{Tereza Mendes}

\affiliation{Instituto de F\'\i sica de S\~ao Carlos, Universidade de S\~ao Paulo, \\
             Caixa Postal 369, 13560-970 S\~ao Carlos, SP, Brazil}

\received{}

\begin{abstract}

Non-perturbative properties of QCD, such as color confinement, are encoded
in the infrared behavior of correlation functions, e.g.\ propagators and vertices.
Various analytic predictions have been suggested for these quantities in
various gauges.
Here we numerically test these predictions using lattice gauge theory.
In particular, we present results for the 2- and 3-point functions
for $SU(2)$ Landau-gauge Yang-Mills theory in three and in four dimensions.
Special attention is paid to systematic finite-volume effects.
The gluon and ghost propagators are also evaluated in the so-called interpolating
gauge (between the Landau and the Coulomb gauge),
in order to study their gauge-dependence.
Finally, we consider these propagators in Landau gauge at finite temperature,
with the aim of understanding the effect of the deconfinement phase transition
on their infrared behavior.
All our results are compatible with the so-called Gribov-Zwanziger confinement
scenario. 


\end{abstract}

\maketitle

\thispagestyle{fancy}
\setcounter{page}{0}


\section{Introduction}

The confinement phenomenon in QCD is a long-range effect and one can expect its
explanation to be encoded in the infrared (IR) behavior of the QCD correlation functions.
Indeed, various confinement scenarios make predictions for this behavior
\cite{Alkofer:2000wg}. Unfortunately, a quantitative determination
of these correlation functions is a very difficult task.

Continuum methods, based e.g.\ on Dyson-Schwinger equations (DSEs)
\cite{Alkofer:2000wg,Fischer,vonSmekal,Zwanziger},
on the renormalization group \cite{Gies} or on effective potentials \cite{Dudal:2005na},
lead to results that seem to agree with the Gribov-Zwanziger confinement scenario
\cite{Zwanziger,Gribov} and with the Kugo-Ojima
scenario \cite{Kugo}.
These methods have the disadvantage of introducing approximations, which in general
cannot be controlled and verified.
On the other hand, numerical studies using lattice gauge theory allow a determination of
QCD correlation functions in the IR limit using approximations that can in principle be
controlled and quantified \cite{Bloch,Sternbeck:2005tk}.
In this case, since one has to work with finite volumes $N^d$, the system considered
has a natural IR cutoff $p \sim 1/(a N)$, making difficult the study of the true IR limit.
(This topic will be discussed in more detail in Section \ref{svolume}.)
Clearly, a close cooperation between continuum and lattice methods can be
a promising approach. In particular, lattice gauge theory allows to check the
approximations usually considered in continuum methods, or at least to
provide constraints for possible Ans\"atze. The most
important test is probably the verification of the approximations employed for the
3-point vertices when solving DSEs.
Results for the three-gluon and the ghost-gluon vertices will be reported
in Section \ref{svertex}.

Most of the studies of QCD correlations functions, both in the continuum and on the
lattice, have been done in Landau gauge. On the other hand, since these quantities
are gauge dependent, it is interesting to evaluate them in other gauges too
(e.g.\ Coulomb gauge \cite{Reinhardt} and maximally Abelian gauge \cite{Bornyakov}).
In principle, the explanation of the confinement mechanism based on
the IR properties of the QCD correlation function can also be gauge-dependent.
Particularly important are the so-called interpolating gauges \cite{Baulieu:1998kx,Capri:2005zj},
since they allow to relate different gauges and possible different confinement mechanisms.
In Section \ref{sinter} we report $3d$ results for the gauge interpolating between the Landau
and the Coulomb gauge.

Finally, a clear understanding of the confinement mechanism could be useful in the
study of the QCD phase diagram and of the deconfining phase transition
\cite{Karsch:2001cy}. In particular, one might expect to observe a modification
of the IR properties of the correlation functions at finite temperature $T$.
As a first step in this direction, the gluon and ghost propagators at finite
$T$ are considered (in the $4d$ Landau gauge) in Section \ref{stemperature}.


\section{Finite-volume effects}\label{svolume}

\begin{figure*}[htbp]
\begin{center}
\includegraphics[width=0.5\linewidth]{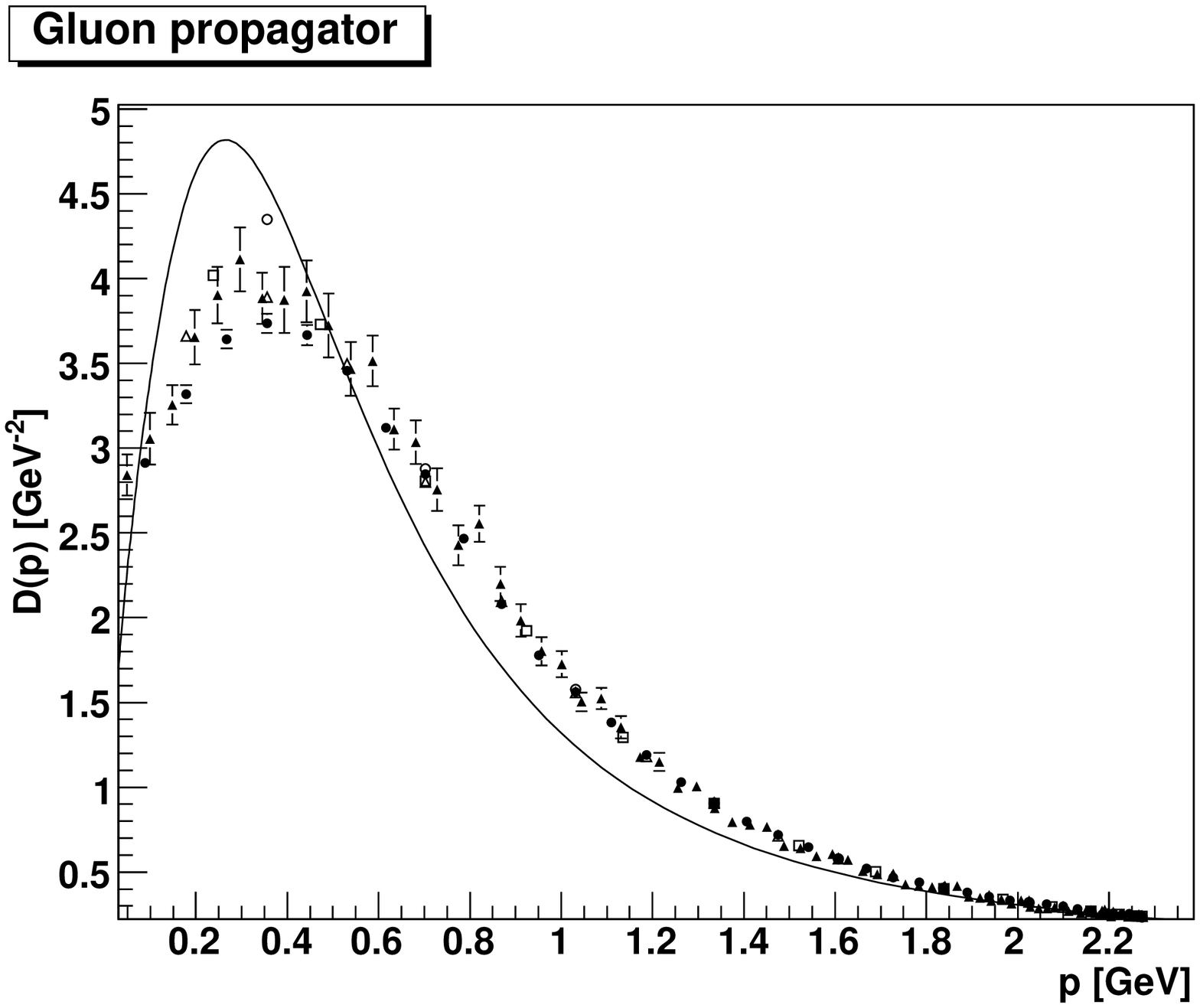}\includegraphics[width=0.5\linewidth]{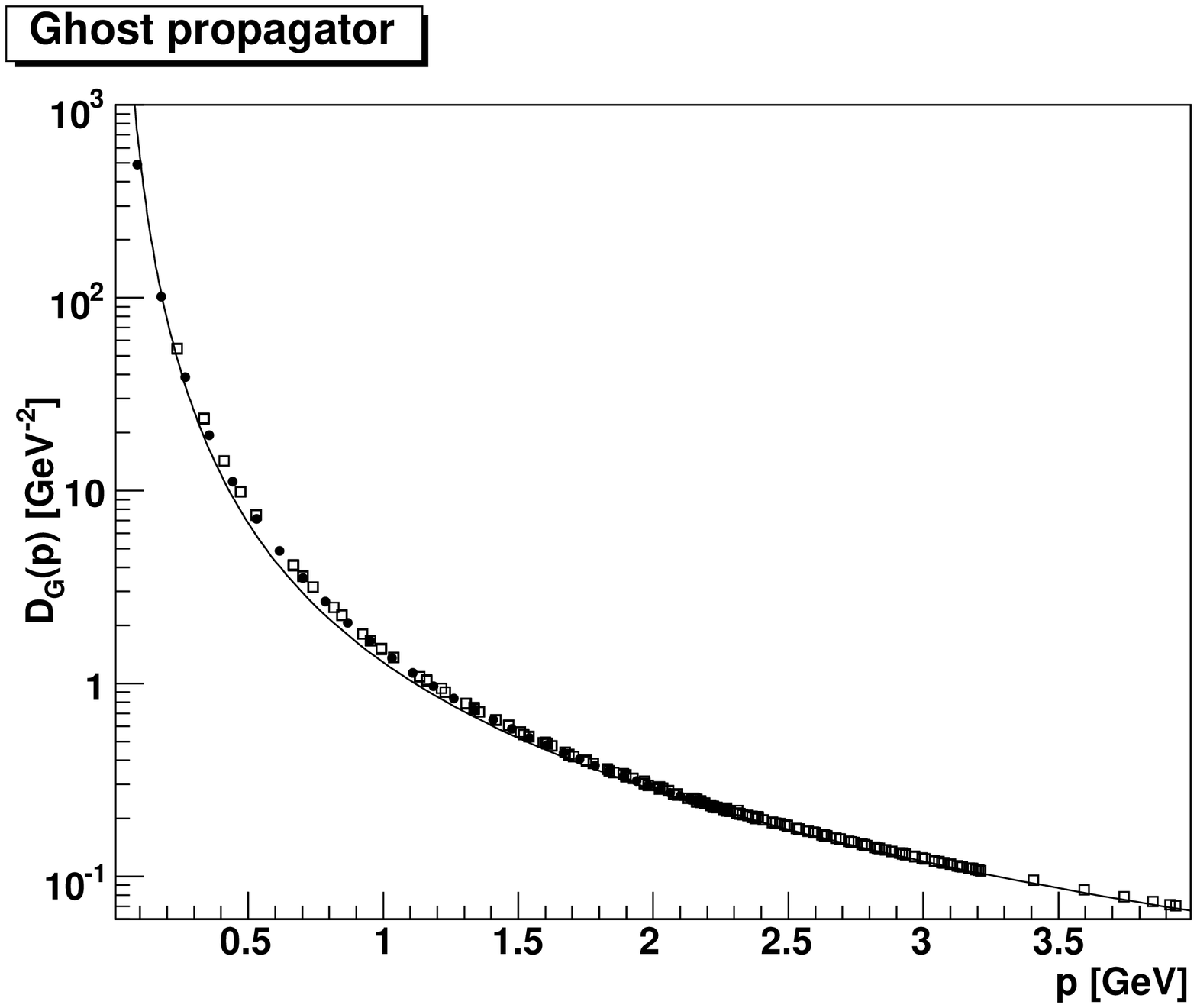}
\caption{Comparison of lattice data and DSE results for gluon $D(p)$ (left) and ghost $D_G(p)$
(right) propagators in the $3d$ case in Landau gauge as a function of the momentum $p$.
The solid line is from DSE results \cite{Maas:2004se}. Lattice simulations
have been done at $\beta=4.2$ considering various lattice volumes. Open circles correspond
to lattice volume $20^3$ $[V\approx(3.5$ fm$)^3]$
\cite{Cucchieri:2006tf}, open squares represent $30^3$ $[V\approx(5.2$ fm$)^3]$ \cite{Cucchieri:2006tf},
open triangles are used for $40^3$ $[V\approx(6.9$ fm$)^3]$ \cite{ftv}, full circles correspond to
$80^3$ $[V\approx(14$ fm$)^3]$ \cite{Cucchieri:2006za} and full triangles represent lattice volume
$140^3$ $[V\approx(24$ fm$)^3]$ \cite{Cucchieri:2003di}. 
\label{fv}}
\end{center}
\end{figure*}

Confinement is induced by long-range correlations. More precisely, the correlation length
should be infinite \cite{Alkofer:2000wg}. Thus, when considering QCD correlation functions in
momentum space, confinement should be related to their IR properties. This is the case for
the Gribov-Zwanziger and the Kugo-Ojima confinement scenarios. Indeed, they predict
a specific IR behavior for the gluon and the ghost propagators in Landau gauge:
the gluon propagator should be IR suppressed --- and it is expected to vanish at zero momentum ---
while the ghost propagator should be IR enhanced compared to the one of a massless particle.

As said above, the IR-cutoff $1/(a N)$ makes it difficult to study the
IR properties of the correlation function and one should pay attention to possible
finite-size effects. Of course, this analysis can be computationally very demanding.
In this case, the consideration of three-dimensional lattices can be useful in order
to access large lattice volumes. Let us note that the Gribov-Zwanziger scenario applies
also to the three-dimensional case \cite{Zwanziger} and that results for the IR behavior
of the propagators using DSEs are available for $d=3$ \cite{Zwanziger,Maas:2004se}.

In Figure \ref{fv} we report the comparison of DSE results with lattice data for the
gluon and the ghost propagators. In the first case, finite-volume effects are evident
for the smallest momenta. In particular, as the volume increases, the propagator becomes more
and more IR suppressed. At the same time, the difference between two different lattices
with a given volume ratio moves further and further into the IR limit with increasing volume.
Thus, momenta of the order of $2\sin(\pi/N)/a \approx 2\pi/(aN)$ are affected by finite-volume effects.
From Figure \ref{fv} one also sees that DSE and lattice results are in
qualitatively agreement.
Let us remark that DSE studies predict \cite{fmps}
that the true IR regime is reached only below a
certain momentum scale $\Lambda_I \lesssim 200$ MeV, i.e.\ on a finite volume
one should have a sufficient number of non-zero momenta
in the range $2\pi/(aN)\ll p\lesssim\Lambda_I\nn$.
With a lattice spacing $a$ of order of 1 GeV$^{-1}$ one needs $N \gg 50$.
Unfortunately, this regime is just reached by the lattices in the $3d$ case but not yet in $4d$.

The ghost propagator is much less affected by finite-volume effects,
even though small effects can be found for the dressing function at small
momenta \cite{ftv}. Moreover, the comparison of the lattice data to DSE results
is quite good (see Fig. \ref{fv}).
One should however recall that the IR enhancement of the ghost propagator seems
to be related to an appropriate sampling of the so-called {\em exceptional configurations}
\cite{Sternbeck:2005tk,Cucchieri:2006tf}. Thus, results obtained considering a
relatively small statistics are likely underestimating this IR enhancement.

We conclude that the smallest non-zero lattice momenta -- of the
order of $2\pi/(aN)$ --- are probably not reliable in lattice
studies and should be treated with great care.
A more thorough discussion of finite-size effects will be presented in \cite{ftv}.


\section{Three-point vertices}\label{svertex}

\begin{figure*}[htbp]
\begin{center}
\includegraphics[width=0.5\linewidth]{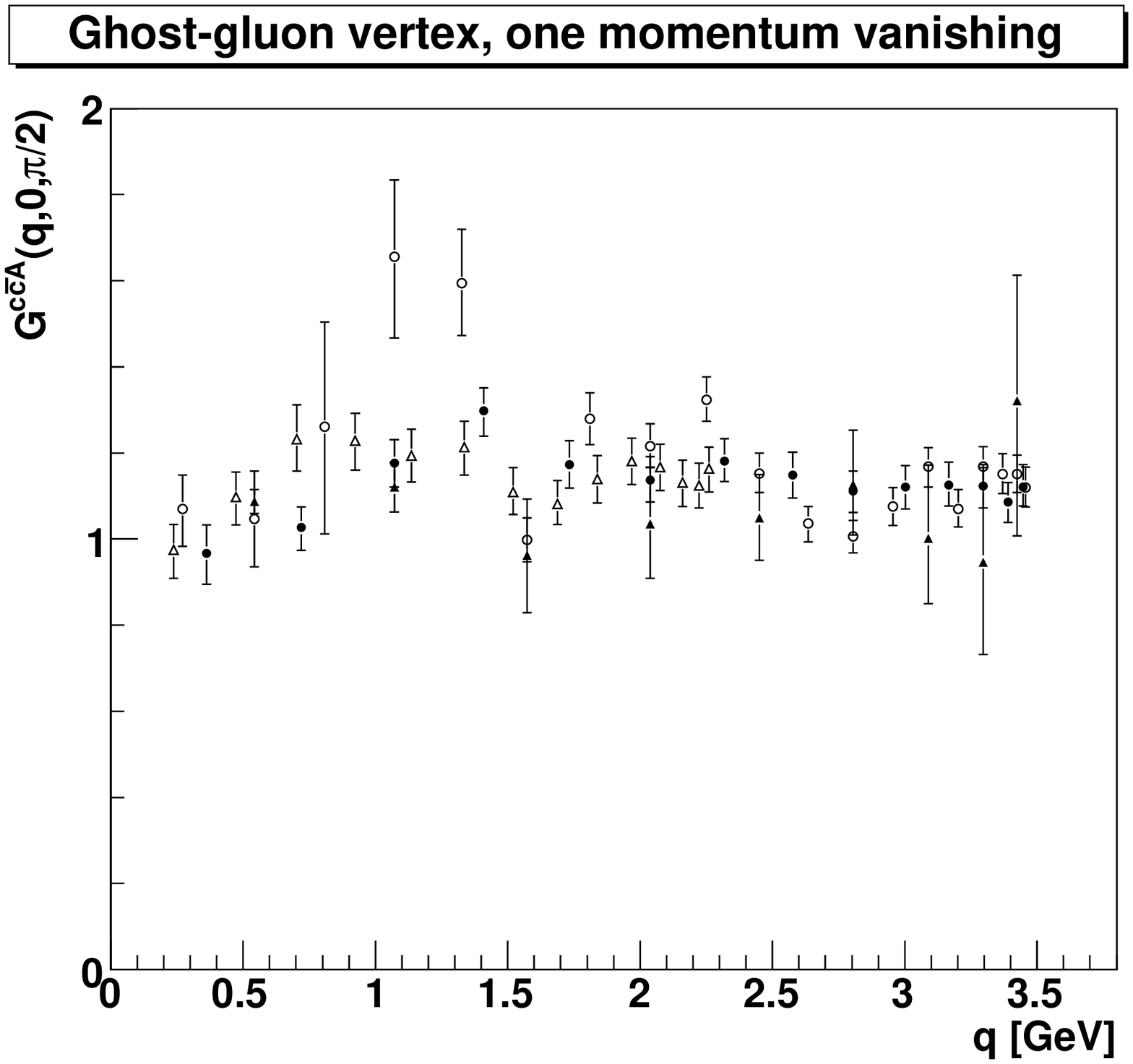}\includegraphics[width=0.5\linewidth]{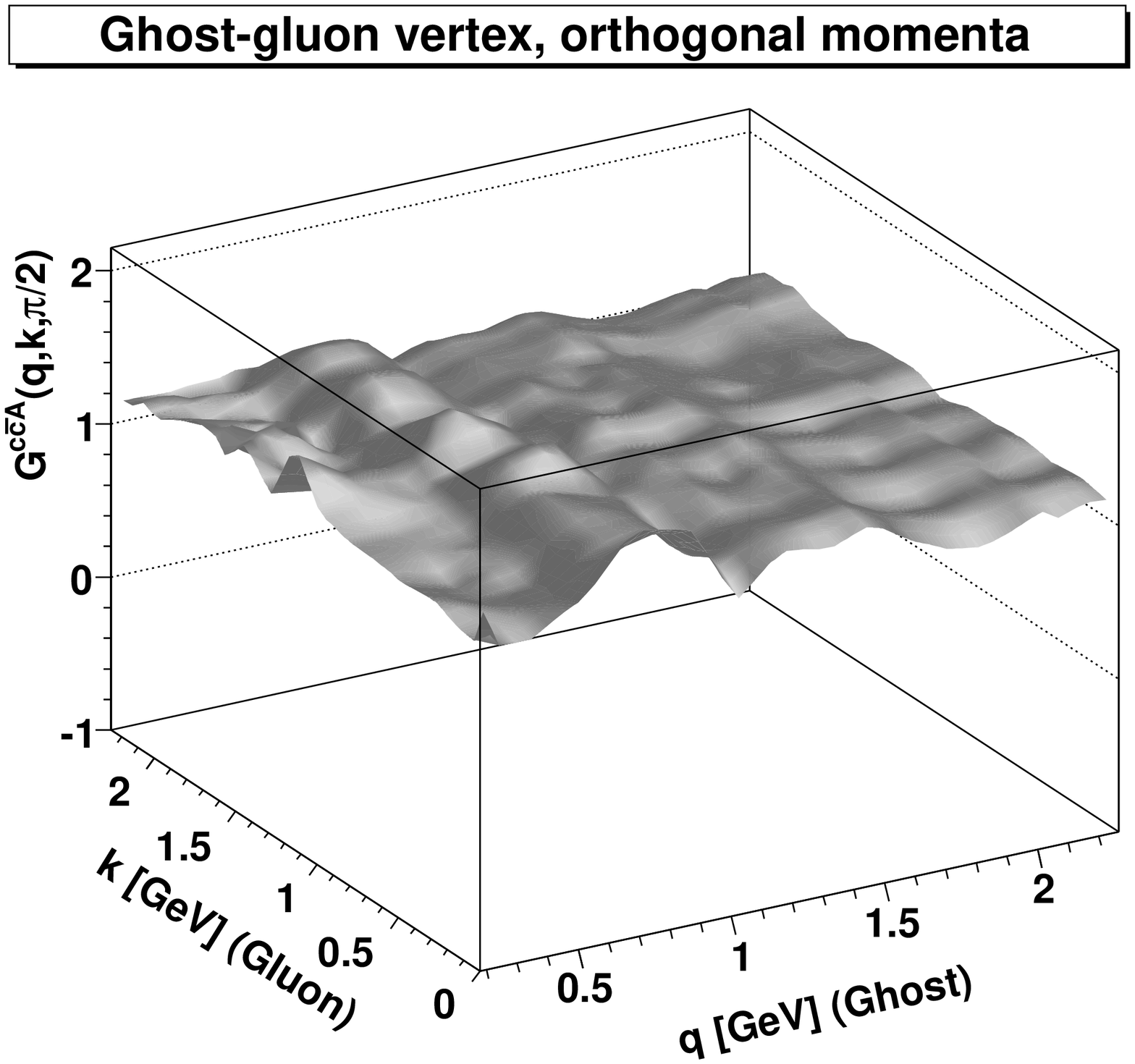}\\
\includegraphics[width=0.5\linewidth]{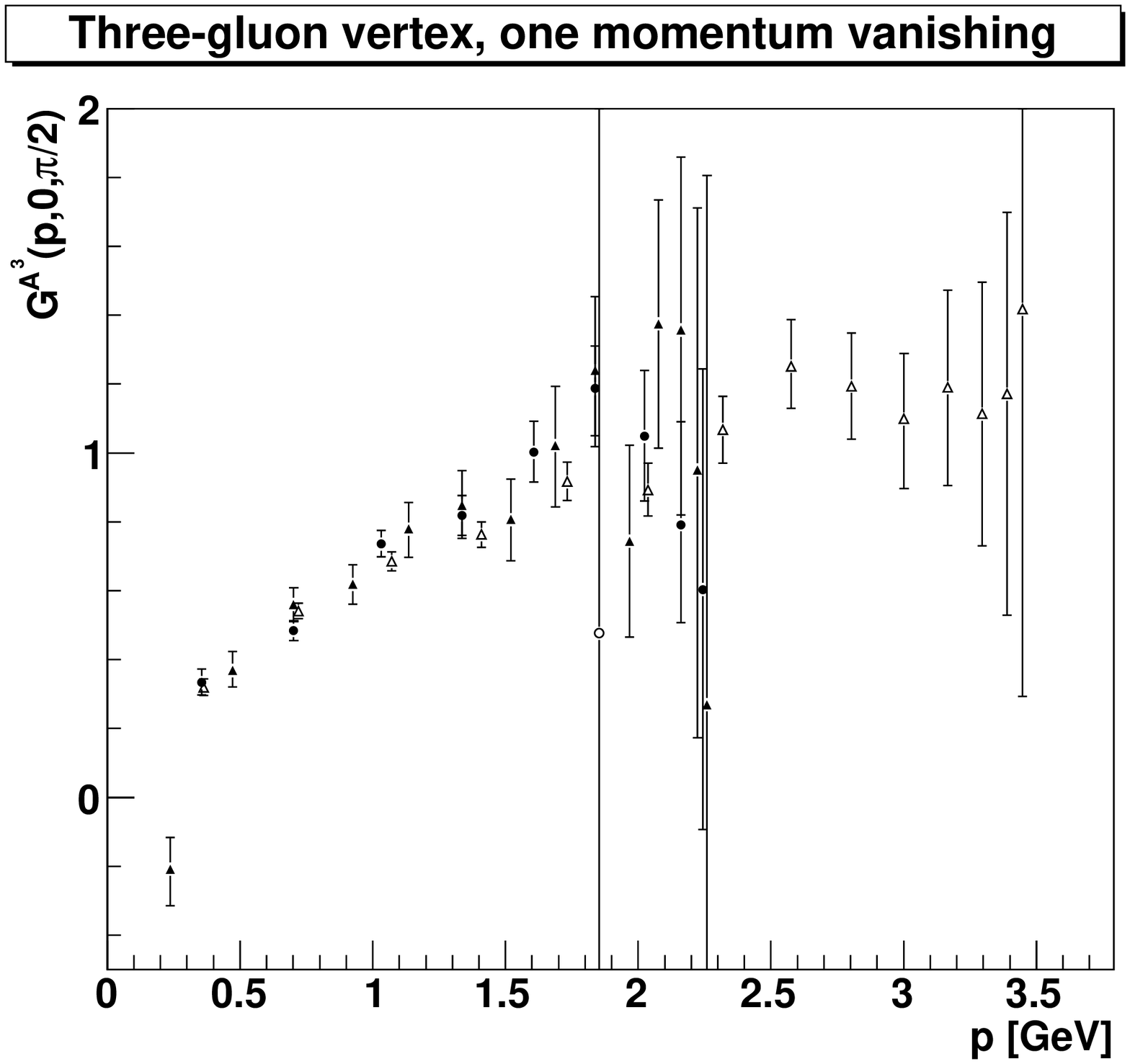}\includegraphics[width=0.5\linewidth]{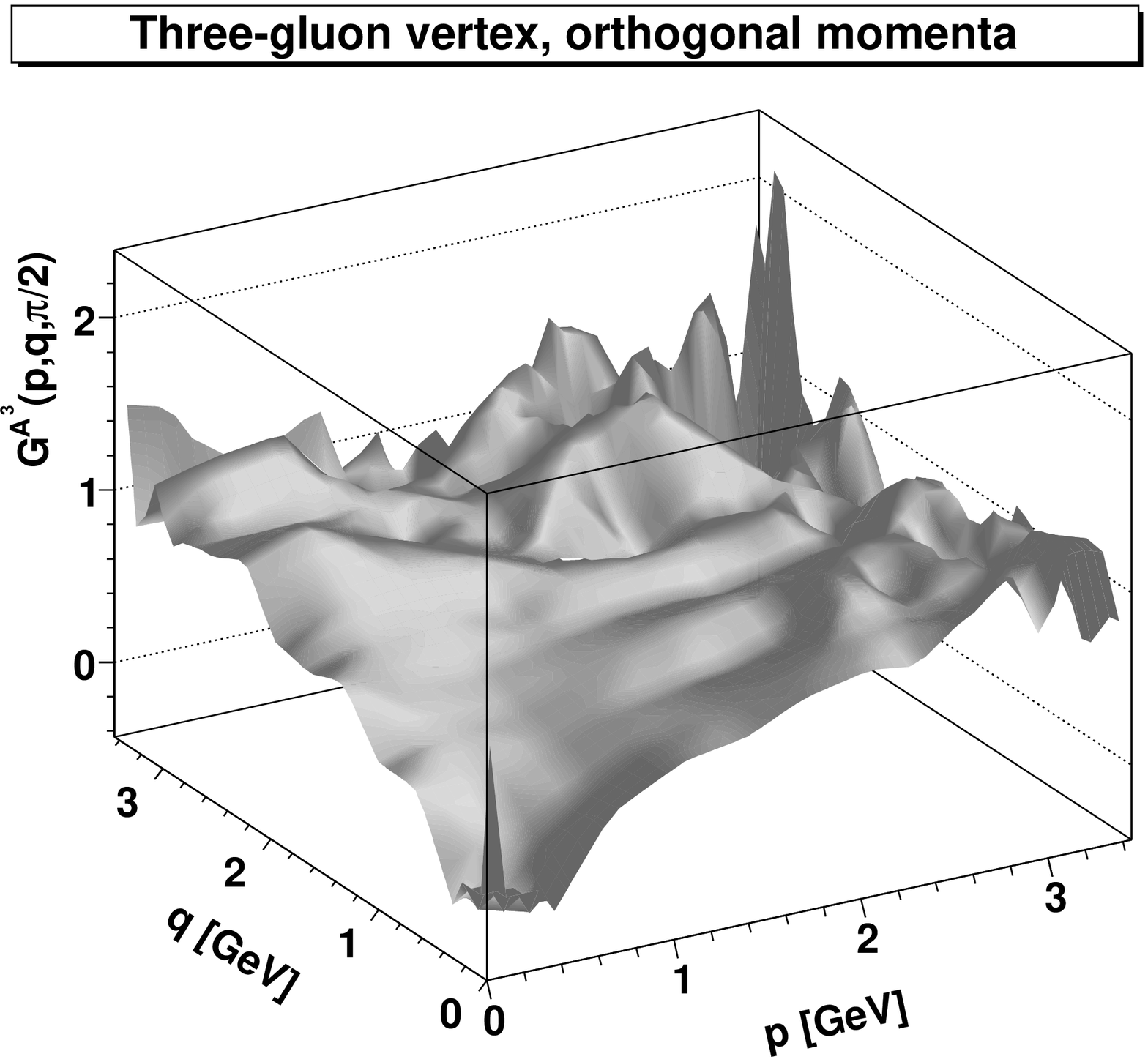}\\
\caption{Top-left panel: ghost-gluon vertex at zero gluon momentum; open triangles are used for $V=30^3$
\cite{vrtx4d} and open circles for $V=30^3$ (in both cases $\beta=4.2$),
full triangles are used for $V=20^3$ \cite{Cucchieri:2006tf} and full circles
for $V=30^3$ (in both cases $\beta=6.0$).
Top-right panel:
ghost-gluon vertex with the gluon and the ghost momenta orthogonal to
each other; the graph interpolates data obtained from a $40^3$ lattice at $\beta=4.2$ \cite{vrtx4d}.
Again, the spike at zero momentum is an artifact, as the vertex function cannot be evaluated there.
Bottom-left panel: 3-gluon vertex with one external momentum vanishing \cite{Cucchieri:2006tf};
full symbols correspond to $\beta=4.2$ and open symbols to $\beta=6.0$; circles are used for
$V=20^3$ and triangles for $V=30^3$.
Bottom-right panel: 3-gluon vertex with two external momenta orthogonal to each other
\cite{Cucchieri:2006tf}; the graph interpolates data obtained from a $30^3$ lattice
at $\beta=4.2$ and at $\beta=6.0$. Note that the spike at zero momentum is an artifact of the
interpolating algorithm, since the vertex cannot be evaluated there.
\label{vrtx3}}
\end{center}
\end{figure*}

\begin{figure*}[htbp]
\begin{center}
\includegraphics[width=0.5\linewidth]{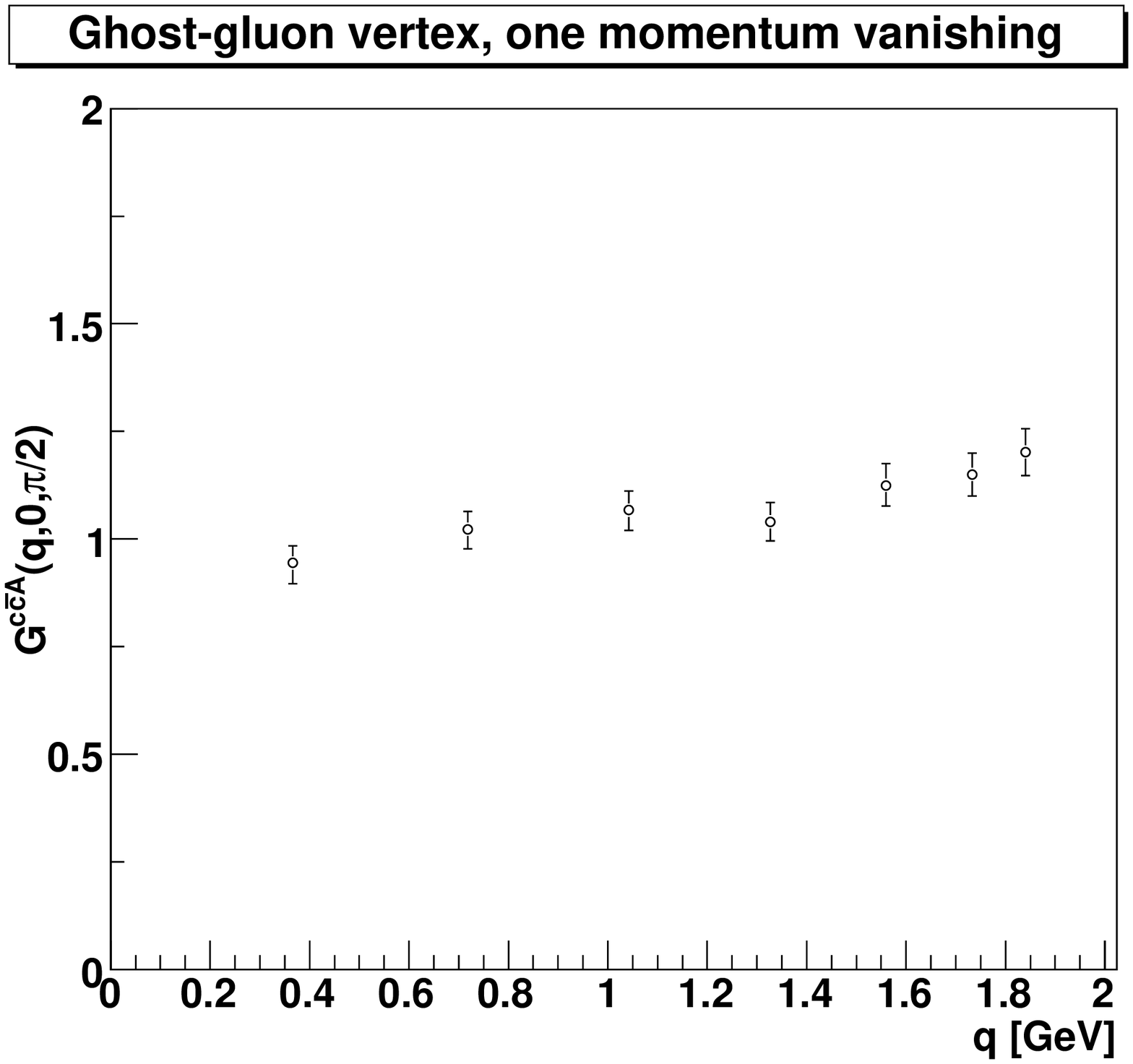}\includegraphics[width=0.5\linewidth]{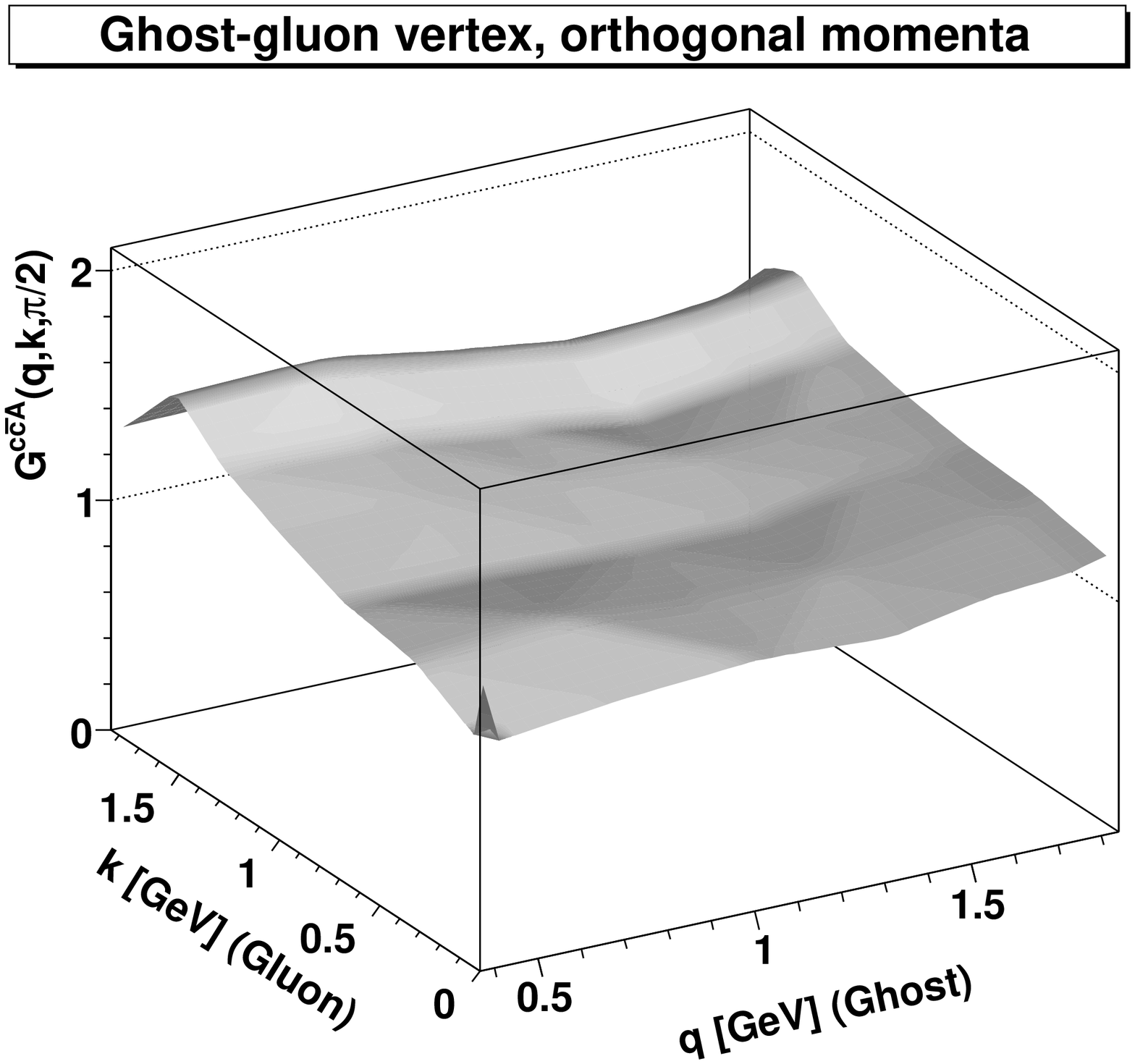}\\
\includegraphics[width=0.5\linewidth]{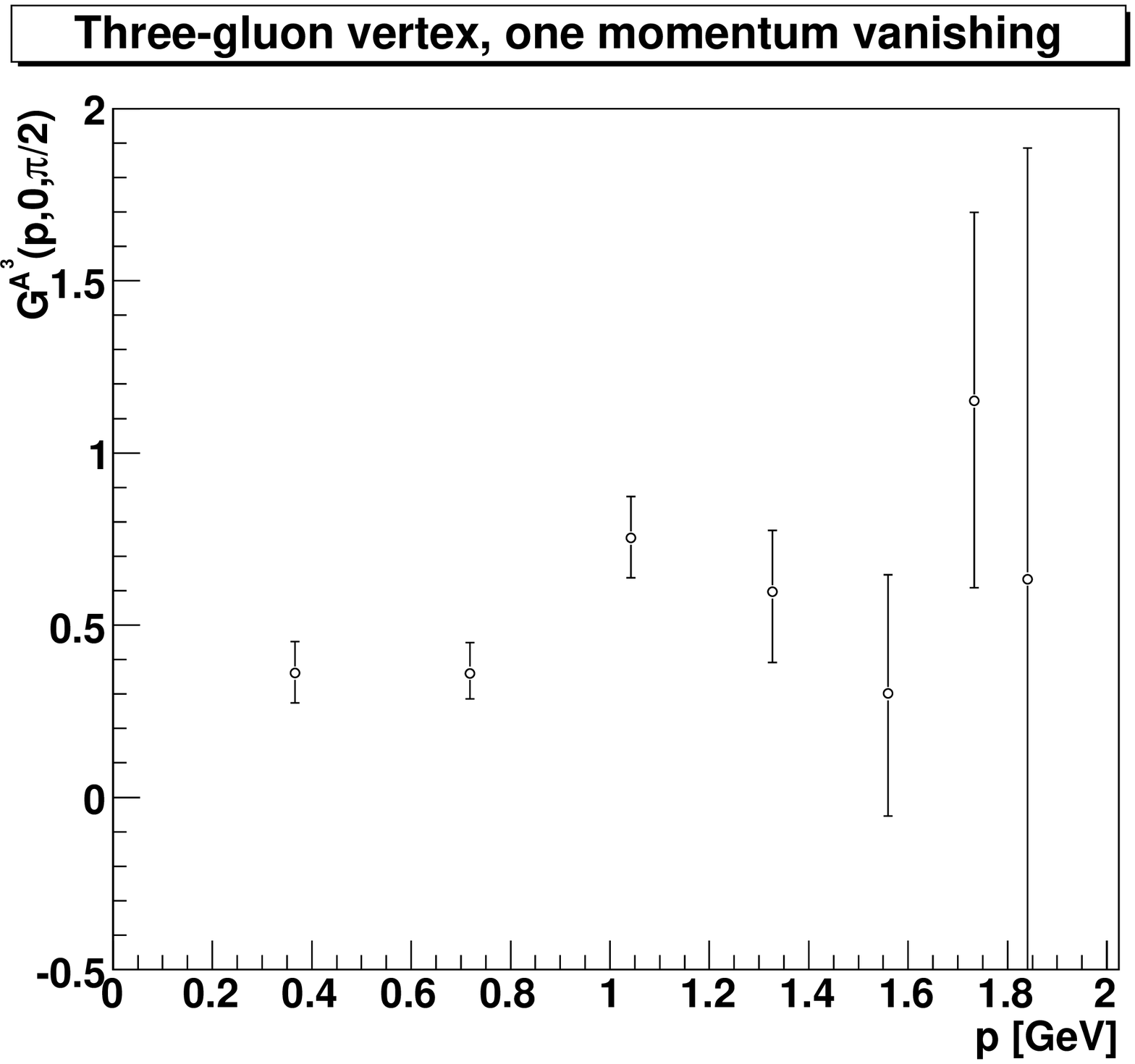}\includegraphics[width=0.5\linewidth]{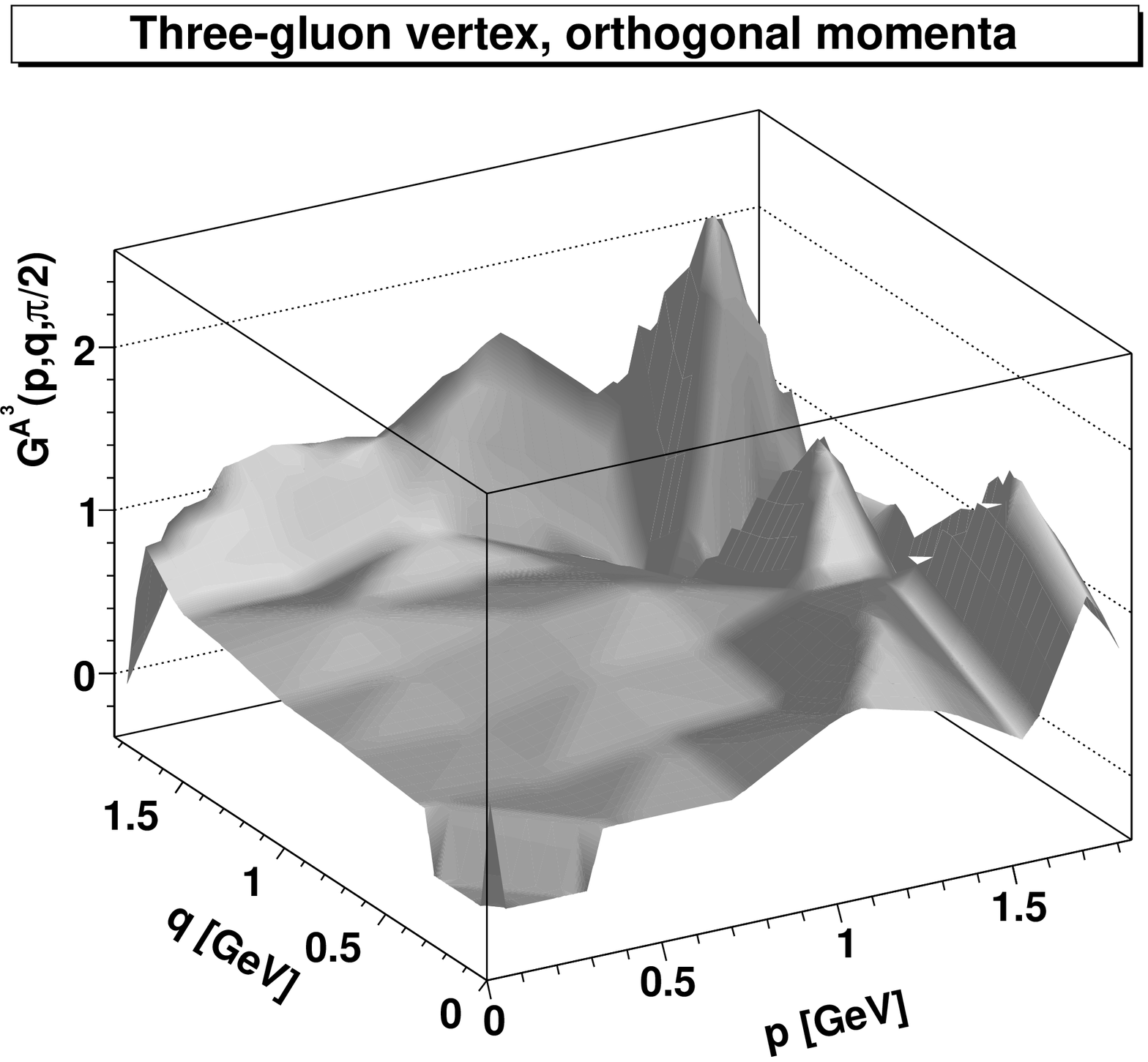}\\
\caption{Same as Figure \ref{vrtx3} but in the four-dimensional case. Here the
lattice volume is $16^4$ and
simulations have been done at $\beta=2.2$ \cite{vrtx4d}.
\label{vrtx4}}
\end{center}
\end{figure*}

Besides the propagators, also the vertices are of great importance in understanding
the IR properties of QCD. In particular,
assumptions on their IR behavior are crucial
inputs for the analytic methods \cite{Alkofer:2000wg,Fischer,vonSmekal}.
At the same time, predictions for their IR behavior have been obtained
using continuum methods \cite{Schleifenbaum:2006bq,Schleifenbaum:2004id,Alkofer:2004it}.
As stressed in the Introduction, lattice gauge theory provides us with the possibility
of verifying these results as well as the assumptions considered in analytic studies.
The simplest of these vertices are the 3-point ones. In Landau-gauge pure
Yang-Mills theory one can consider
the ghost-gluon vertex and the three-gluon vertex. Kinematically these vertices are much
more complicated than the propagators, since they depend on three independent kinematic variables.
In addition, they have a complex tensor structure (especially the three-gluon vertex).

In order to simplify the presentation, we consider here only one specific kinematic configuration,
namely the one in which two of the incoming momenta are orthogonal with respect to each
other (see \cite{Cucchieri:2006tf,vrtx4d}
for details and for other possible kinematic configurations). Furthermore, instead of the various tensor
elements we evaluate the quantity \cite{Cucchieri:2006tf}
\be
G=\frac{\Gamma^{\tl\indexsep abc}G^{abc}}{
   \Gamma^{\tl\indexsep abc}D^{ad}D^{be}D^{cf}\Gamma^{\tl\indexsep def}}\label{vdef}.
\ee
\no
Here the indices $a, \ldots, f$ are generic multi-indices, encompassing field-type, Lorentz
and color indices. Also, $D^{ab}$ are the propagators of the fields, $G^{abc}$ represent
the full Green's functions and $\Gamma^{\tl\indexsep abc}$ are the corresponding
tree-level vertices. This quantity is defined such that it becomes equal to one
if the full and the tree-level vertex coincide.
Furthermore, the ratio \pref{vdef} corresponds to the kernel of a
one-loop not-amputated self-energy diagram in DSEs over the same kernel with
full vertices replaced by the bare ones. Thus, the quantity $G$ is a direct measure of
how good the approximations in DSE calculations are when the bare vertices are used.

Results for the quantity \pref{vdef} in three and in four dimensions for both vertices are shown
in Figures \ref{vrtx3} and \ref{vrtx4}, respectively.
The ghost-gluon vertex is essentially constant for all momenta studied here as well
as for other kinematic configurations \cite{Cucchieri:2006tf,vrtx4d,Cucchieri:2004sq},
both in $3d$ and in $4d$.
In particular, there is no sign of enhancement or suppression even for the smallest momenta
available. This is in agreement with results obtained using continuum methods
\cite{Schleifenbaum:2004id,Alkofer:2004it}. In addition, the constancy of this vertex
confirms the most important assumption made in DSE studies, i.e.\ the use of a bare ghost-gluon
vertex \cite{Alkofer:2000wg,Fischer,Maas:2004se,Alkofer:2004it}.

The evaluation of three-gluon vertex is numerically much more demanding than that of 
the ghost-gluon vertex.
In particular, very large statistics are required for its investigation \cite{Cucchieri:2006tf,vrtx4d},
especially in the three-dimensional case. Our results show that this vertex is suppressed at
intermediate momenta, especially for $d=3$. On the other hand, with our data,
it is difficult to understand completely its precise IR behavior.
Let us note that, if this suppression were to be confirmed by studies using larger lattices, 
this result would contradict predictions obtained from DSE studies
\cite{Schleifenbaum:2006bq,Alkofer:2004it}. Nevertheless, it would not invalidate the
assumptions usually made in the continuum calculations of the propagators
\cite{Alkofer:2000wg,Fischer,Maas:2004se,Alkofer:2004it}.


\begin{figure}[htbp]
\begin{center}
\includegraphics[width=\linewidth]{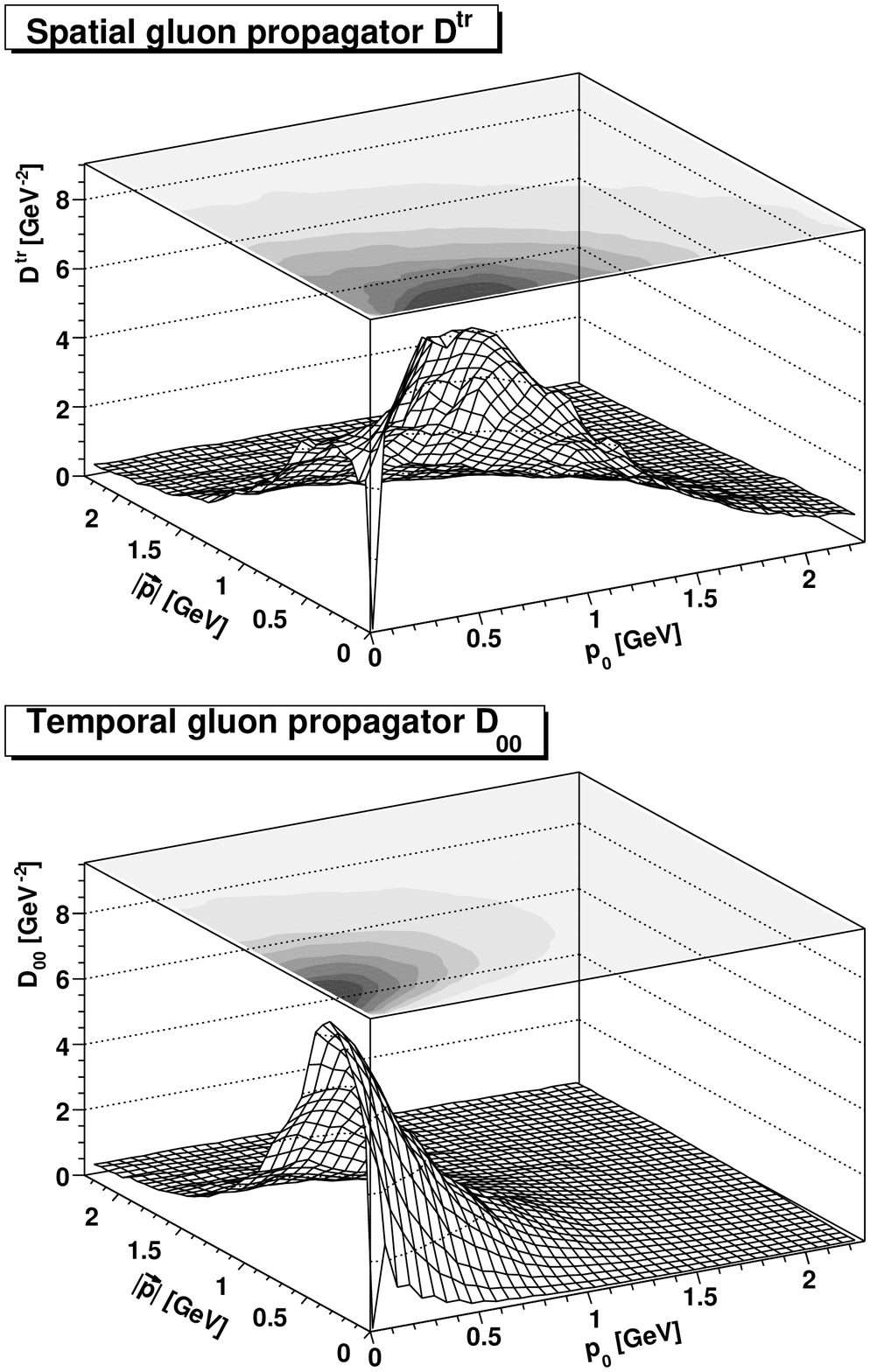}
\caption{The spatial (top) and temporal (bottom) gluon propagator
in the interpolating gauge \pref{ig} at $\lambda=1/2$ as a function of
the temporal $p_0$ and spatial $|\vec{p}|$ momenta, from a $60^3$
lattice at $\beta=4.2$ \cite{intg}.
Results have been interpolated; the spike at zero is an artifact of the
missing point there.
\label{intgp}}
\end{center}
\end{figure}

\begin{figure}[htbp]
\begin{center}
\includegraphics[width=\linewidth]{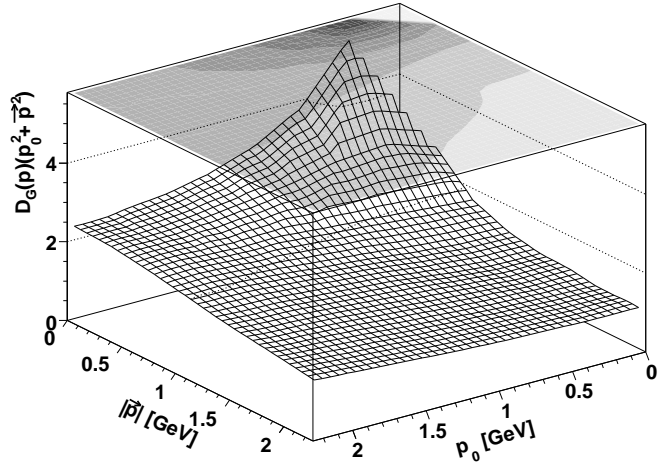}
\caption{The ghost dressing function $D_G(p)(p_0^2+\vec p^2)$
in the interpolating gauge \pref{ig} at $\lambda=1/2$ as a function of the temporal
$p_0$ and spatial $|\vec{p}|$ momenta, from a $20^3$ lattice at
$\beta=4.2$ \cite{intg}. Numerical data have been interpolated.
\label{intghp}}
\end{center}
\end{figure}

\section{Interpolating gauges}\label{sinter}

All results presented so far have been obtained in the (minimal) Landau gauge.
As said in the Introduction, it is eventually desirable to understand confinement
in a gauge-independent way, if this is possible. Thus, the IR properties of
the QCD correlation functions in other gauges are also interesting.
Among other possible gauge-fixing conditions, the so-called interpolating gauges
allow to interpolate smoothly between specific gauge conditions.
One example is given by the gauge condition \cite{Baulieu:1998kx,Cucchieri:1998ta}
\be
\lambda \,\pd_0 A_0^b+\vec{\pd}\vec{A}^b=0.\label{ig}
\ee
\no
This gauge interpolates between the Landau gauge ($\lambda=1$) and the Coulomb gauge ($\lambda=0$),
even though the Coulomb gauge limit itself is likely not to be a smooth one \cite{Fischer:2005qe}.
Nonetheless, this gauge allows one to study the Gribov-Zwanziger scenario when moving away
from the Landau gauge.
Since for $\lambda \neq 1$ the $O(4)$-invariance is explicitly broken, it is natural to consider
two scalar gluon propagators \cite{Fischer:2005qe}, i.e.\ the time-time component
$D_{00}^{aa}/(N_c^2-1)$ and the $3$-dimensional transverse
one $D^{\mathrm{tr}}=D^{aa}_{ii}/((d-1)(N_c^2-1))$.
In addition, the propagators depend independently on the temporal momentum $p_0$
and on the spatial momentum $|\vec{p}|$.

Results at $\lambda=1/2$ are shown in Figure \ref{intgp} for these two gluon propagators
and in Figure \ref{intghp} for the scalar ghost dressing function $D_G(p)(p_0^2+\vec p^2)$.
Note that $D_{00}(p_0,0)$ vanishes exactly for any $p_0$ due to the gauge
condition, even though (due to technical reasons)
this is not clear from Figure \ref{intgp}.
Both gluon propagators are found to be IR suppressed when considered as a function
of $p_0^2+\vec p^2$. Also, both propagators exhibit a clear maximum.
Furthermore, both propagators are not suppressed as long as either $p_0$ or $|\vec{p}|$
are not sufficiently small.
At the same time, the ghost propagator is IR enhanced. The difference
in behavior at large pure-temporal and at large pure-spatial 
momenta is a direct consequence of how
$\lambda$ appears in the tree-level propagators.
Thus, the IR behavior of these propagators is (qualitatively) the same
as that found in Landau gauge. This is in agreement with results
from DSE studies \cite{Fischer:2005qe}. On the other hand, since the
limit $\lambda \to 1$ is smooth, one wouldn't expect a strong deviation from the Landau
behavior for a value $\lambda = 1/2$. Results for the non-smooth limit $\lambda\to 0$
will be presented elsewhere \cite{intg}.


\begin{figure}[htbp]
\begin{center}
\includegraphics[width=\linewidth]{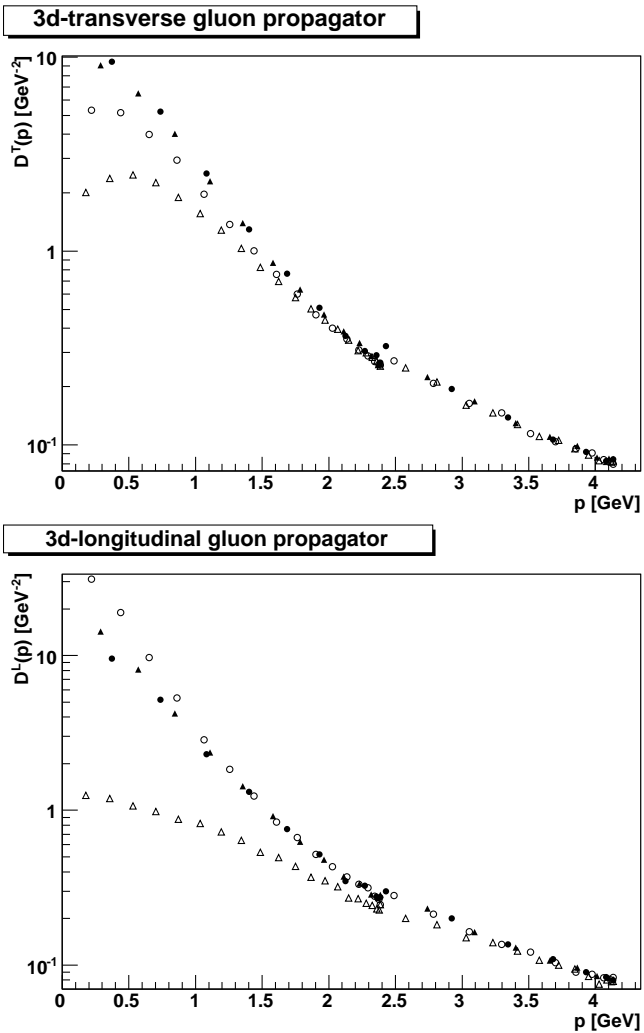}
\caption{The spatially transverse (top) and spatially longitudinal (bottom)
gluon propagators at various temperatures \cite{ftv}. Full circles are at $T=0$
on a $20^4$ lattice, full triangles are at $T\approx 149$ MeV on a $10\times 26^3$ lattice,
open circles are at $T\approx 298$ MeV on a $4\times 34^3$ lattice, and open triangles are
at $T\approx 597$ MeV on a $2\times 42^3$ lattice. All results have been obtained using 
$\beta=2.3$.
\label{ftgp}}
\end{center}
\end{figure}

\begin{figure}[htbp]
\begin{center}
\includegraphics[width=\linewidth]{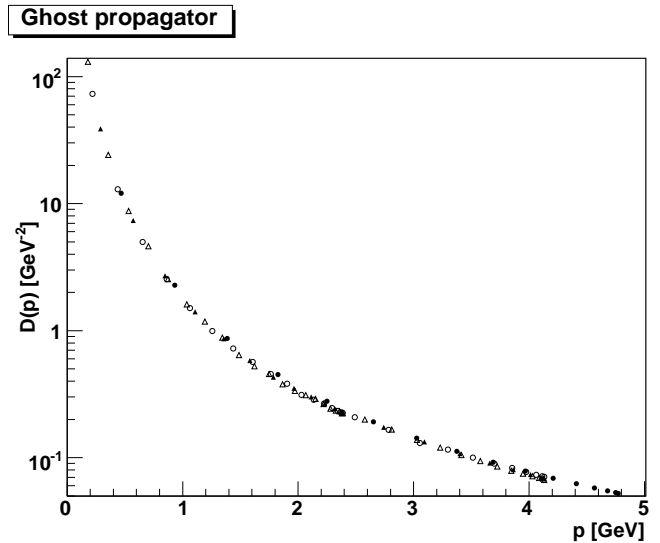}
\caption{Same as figure \ref{ftgp}, but for the ghost propagator. Also the symbols are
the same except for full circles, which refer to a $32^4$ lattice at $\beta=2.3$
($T=0$).
\label{ftghp}}
\end{center}
\end{figure}

\section{Finite temperature}\label{stemperature}

Yang-Mills theory undergoes a deconfining phase transition at finite
temperature \cite{Karsch:2001cy} and one expects this transition to
change the confining properties of the theory. At the same time,
if confinement is manifest in the IR properties of the QCD propagators,
these might also change across the phase transition.
At finite temperature, the gluon propagator can no longer be described by
a single tensor structure and one needs two independent structures \cite{Kapusta:tk}:
\be
D_\mn^{ab}=P_\mn^T D^{Tab}+P_\mn^L D^{Lab}.\nn
\ee 
\no
Here, $P_\mn^T$ is purely spatial and transverse while $P_\mn^L = P_\mn-P_\mn^T$,
i.e.\ $P_\mn^L$ complements $P_\mn^T$ to a four-dimensional-transverse $O(4)$-invariant
tensor-structure $P_\mn=\delta_\mn-p_\mu p_\nu/p^2$.
In addition, $P_\mn^L$ is longitudinal in the spatial sub-space.
Furthermore, all functions depend on the spatial momentum $|\vec{p}|$
and on the energy $p_0$ separately. The latter is discrete with values
$p_0 = 2\pi T n$, where $n$ is an integer. Infrared or soft modes are those
with zero energy, i.e.\ $n = 0$.

Results for the soft modes of the two gluon propagators and of the (scalar) ghost propagator
are shown in Figures \ref{ftgp} and \ref{ftghp} for temperatures above and below the phase
transition. Clearly, as expected, all propagators coincide with the zero-temperature perturbative tail
for momenta $p$ large compared to the temperature $T$. At the same time, each of the
three propagators shows a different temperature dependence at low momenta.
In particular, the $3d$-transverse propagator $D^T(p)$
decreases smoothly (in the IR) with increasing temperature. At the same time,
for the highest temperature considered here ($T \approx 597$ MeV),
one sees a clear IR suppression with a maximum for $p \approx 600$ MeV.
Let us stress that, while at $T = 0$ one does not see a decreasing propagator
$D^{T}(p)$ even for a lattice volume of $52^4$ \cite{Cucchieri:2006xi},
for this temperature above the
phase transition the IR suppression is already manifest for a spatial volume
of $42^3$. This suggests that, at high temperature, either the infrared suppression
is stronger or that finite-volume effects are smaller, compared to the
zero-temperature case.
In the case of the $3d$-longitudinal propagator $D^L(p)$ we observe
an increase as the temperature is turned on starting from $T=0$.
However, the propagator decreases (in the IR region) when going from
the second-highest to the highest temperature. A possibility would be 
a propagator increasing (respectively decreasing) when the temperature rises
for temperatures below (respectively above) the phase transition.
Let us note that $D^L(p)$ does not seem to be IR suppressed, even though
in the IR limit it does not grow as strongly as the propagator of a massless particle \cite{ftv}.
Finally, the ghost propagator does not show any visible temperature dependence.
Let us also stress that these results should be taken with care,
especially when considering the $3d$-longitudinal
gluon propagator $D^{L}(p)$, since we find strong volume and discretization effects\cite{ftv}.

These results are in agreement with previous lattice \cite{Cucchieri}
and continuum results \cite{Zahed} and they can be interpreted
considering a Gribov-Zwanziger-type scenario \cite{ftv}.
In particular, they suggest that this confinement scenario
prevails at all temperatures, i.e.\
not only for small temperature, but also in the high-temperature phase.
 

\section{Summary}

Summarizing, we have presented results for gluon and ghost propagators
considering the 3- and the 4-dimensional cases for various gauges at
zero temperature.
In all cases our results are in qualitatively agreement with the
Kugo-Ojima and with the Gribov-Zwanziger scenarios.
Moreover, an investigation of the propagators in the interpolating gauge
\pref{ig} suggests that the Gribov-Zwanziger scenario is stable when moving away from
the Landau gauge condition.

At zero temperature, in Landau gauge, we have also
evaluated the ghost-gluon and the 3-gluon vertices.
This study provides additional hints and constraints for the assumptions
usually considered in continuum studies.

Finally, we have presented results for the two propagators at finite temperature
around the deconfining phase transition.
These results fit into an extension of the Gribov-Zwanziger
scenario at finite temperature \cite{ftv,Cucchieri,Zahed}.

In all cases considered, the analysis of finite-volume effects
should be done very carefully, especially for the gluon propagator.
Thus, a complete understanding of the IR behavior of the QCD
correlation functions and a confirmation (or disproof) of the
available confinement scenarios will be obtained only when
extending these studies to larger lattices.


\acknowledgments

A.\ M.\ was supported by the DFG under grant number MA 3935/1-1.
A.\ C.\ and T.\ M.\ were supported by FAPESP (under grant \# 00/ 05047-5)
and by CNPq.



\end{document}